\documentclass[aps,floatfix,showpacs,prb]{revtex4-2}
\usepackage{graphicx,amsfonts,amssymb,enumerate,color}
\usepackage{xcolor}
\begin{document}
%%%%%%%%%%%%%%%%%%%%%%%%%%%%%%%%%%%%%%%%%%%%%%%%%%%%%%%%%%%%%%%%%%%%%%%%%%%%%%%%%%
\title{Entropic selection of magnetization in a frustrated 2D magnetic model}

\author{Anuradha Jagannathan}
\affiliation{Laboratoire de Physique des Solides, 
Universit\'{e} Paris-Saclay, 91400 Orsay, France}

\author{Thierry Jolicoeur}
\affiliation{Universit\'e Paris-Saclay, CNRS, CEA, 
Institut de Physique Th\'eorique, France}

\date{January 25th, 2025}

%%%%%%%%%%%%%%%%%%%%%%%%%%%%%%%%%%%%%%%%%%%%%%%%%%%%%%%%%%%%%%%%%%%%%%%%%%%%%%%%
\begin{abstract}
We discuss the magnetic ground state and properties of a frustrated
two-dimensional classical Heisenberg model of interacting hexagonal clusters of spins. The energy of the ground states is found exactly for arbitrary values of $J_1$ (intra-cluster couplings) and $J_2$ (inter-cluster couplings). Our main results concern a frustrated region of the phase diagram, where we show that the set of ground states 
has a degeneracy larger than that due to global rotation symmetry. Furthermore, the ground state manifold
does not have a fixed total magnetization~: there is a range of allowed values.
At finite temperature, our Monte-Carlo simulations show that the entropy
selects the most probable value of the total magnetization, while
 the histogram of the Monte-Carlo time series is 
non-trivial. This model is a first step towards modelling properties of a class of frustrated magnetic structures composed of coupled spin clusters.
\end{abstract}


\maketitle
%%%%%%%%%%%%%%%%%%%%%%%%%%%%%%%%%%%%%%%%%%%%%%%%%%%%%%%%%%%%%%%%%%%%%%%%%%%%%%%%%%%

\section{Introduction}
Frustration occurs in many different magnetic systems, and can lead to rich phase diagrams,
with or without long range magnetic order, depending on the lattice and interactions.  
In this paper, we introduce a 2D model of clusters of spins which are placed on vertices 
of a periodic lattice. The competition between intra-cluster and inter-cluster couplings 
can result in frustration and, as we will show, some novel phenomena at zero and at finite 
temperature. The model itself takes its motivation from a class of experimental 
three-dimensional magnets containing magnetic rare earths in a metallic matrix. The rare 
earth spins primarily sit on icosahedral clusters, which are coupled by RKKY interactions. 
The properties of such frustrated magnets have been explored by experiments~\cite{tamura,labib} 
for a variety of periodic and quasiperiodic structures, and numerical simulations have been 
carried out~\cite{matsuo,sugimoto}, however there have been no systematic theoretical studies. 
Thus, one of the future goals of our study, although not the focus of the present paper, is 
to elucidate the phase diagram of such frustrated systems. As a starting point towards this goal, 
we consider spin cluster models based on periodic approximants of square-triangle tilings. 
The family of square triangle tilings is very diverse, and enters for example in the description 
of Frank-Kasper phases and of dodecagonal quasicrystals~\cite{ungarzeng,marianne}.
We recall that frustrated classical spin systems have been extensively studied. 
Two examples of these are the triangular and Kagome antiferromagnets, with nearest-neighbor 
Heisenberg couplings between spins. The former has a unique ground state (up to global rotations), 
whereas in the latter, there is a macroscopic number of degenerate ground states~\cite{Chandra90}. 
It was shown that the degeneracy can be lifted at finite temperature by entropic effects, 
and this phenomenon is called 
``order-by-disorder"~\cite{Reimers91,Harris92,Chalker92,Chubukov92,Coffey92,Huse92,Reimers93,Ritchey93,Berlinsky94,Henley95}. In the Kagome lattice the degeneracy of the ensemble of ground states can be continuous, 
due to out-of-plane fluctuations around a planar state, for example, or discretely countable. 

\begin{figure}
\centering
\includegraphics[angle=90,width=0.3\linewidth]{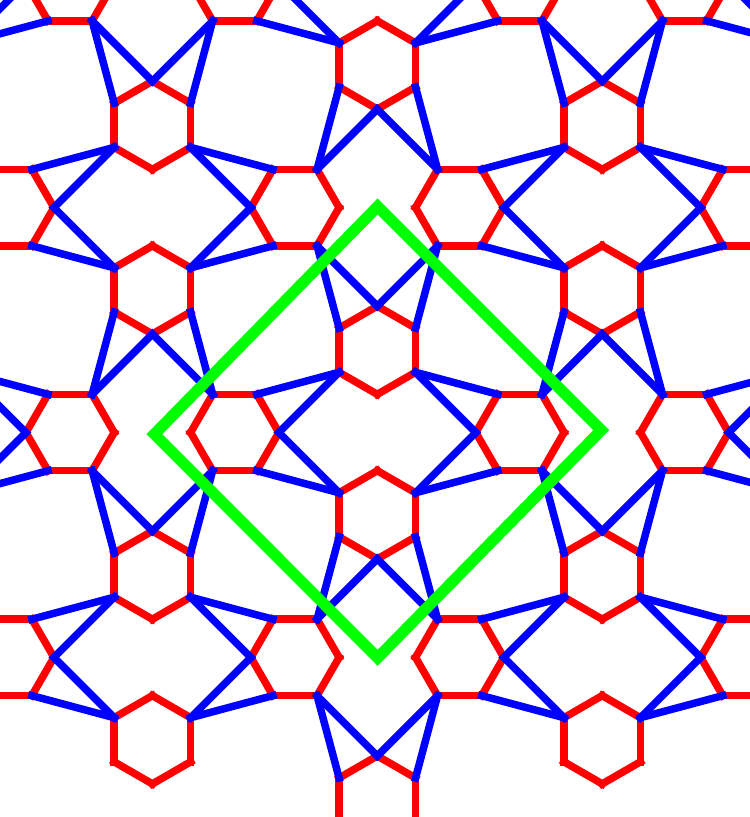}
\caption{The H-$\sigma$ lattice : the unit cell (outlined in green) consists of
four hexagons with $J_1$ bonds (in red) inside hexagons
and $J_2$ bonds relating hexagons (in blue). }
\label{figHsigmalattice}
\end{figure}

Among the frustrated lattice we have studied stands out a very special model
that we consider in this paper. It is defined by placing hexagonal clusters on vertices of the so-called 
sigma lattice, according to a rule developed by Schlottman for dodecagonal quasicrystals that is 
described in ref.(\onlinecite{hermisson}). We have termed the resulting structure the Hex-on-sigma 
(H-$\sigma$) lattice~\footnote{Interestingly, a different structure obtained by putting hexagonal 
clusters on the square lattice turns out to be closely related to the well-known Shastry-Sutherland 
model (Physica 108B (1981) 1069)}. Fig.\ref{figHsigmalattice} shows the unit cell (outlined in green) 
of this structure, which has 24 spins. There are four hexagons per unit cell, having two 
different orientations. 
We study the isotropic classicalHeisenberg model on this lattice with intra-cluster
Heisenberg exchange
couplings $J_1$ (red) and inter-cluster couplings $J_2$ (blue). It can be seen that there are 
triangles in this structure, these give rise to frustration when bonds are antiferromagnetic.  
This paper describes the phase diagram in the $J_1$-$J_2$ plane for this system, with special focus on 
the non-trivial frustrated region which is $0<J_2<J_1$

In this range of couplings the ground state has an extensive degeneracy beyond the global
rotation of spins. The ground state manifold does not have a fixed magnetization.
Indeed we find that there is a finite range of values for the total magnetization
which does not shrink when the system becomes large. At finite temperatures
Monte-carlo studies shows that there is selection of a preferred value
of the magnetization with a non-trivial distribution which does not beacomes Gaussian
for large system sizes.
It is the entropy that select the most probable spin configuration.
 
 The organization of the paper is as follows.
In section II we discuss several approximants for the description
 of quasicrystals and introduce the $\sigma$ lattice as well as its derivative of
 interest~:
 the H-$\sigma$ lattice.
 Section III introduces the Heisenberg model on the H-$\sigma$ lattice
 and discuss the family of ground states.
 Finite temperature properties are obtained from classical metropolis Monte-Carlo
 simulations described in section IV. The effect of an applied magnetic field
 is discussed in section V. An additional exchange coupling
 is introduced and studied in section VI. Our conclusions are given in
 section VII.
 
%%%%%%%%%%%%%%%%%%%%%%%%%%%%%%%%%%%%%%%%%%%%%%%%%%%%%%%%%%%%%%%%%%%%%%%%%%%%%%%%%%%

%%%%%%%%%%%%%%%%%%%%%%%%%%%%%%%%%%%%%%%%%%%%%%%%%%%%%%%%%%%%%%%%%%%%%%%%%%%%%%%%%
\section{Structure of the H-$\sigma$ lattice}
\begin{figure}
\centering
\includegraphics[width=0.37\textwidth]{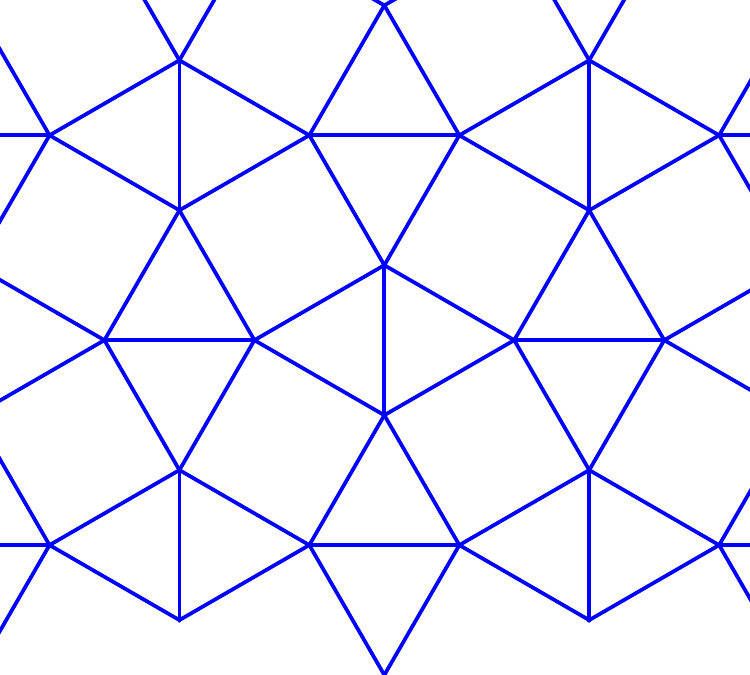} \hskip 1.4cm
\includegraphics[angle=90,width=0.3\linewidth]{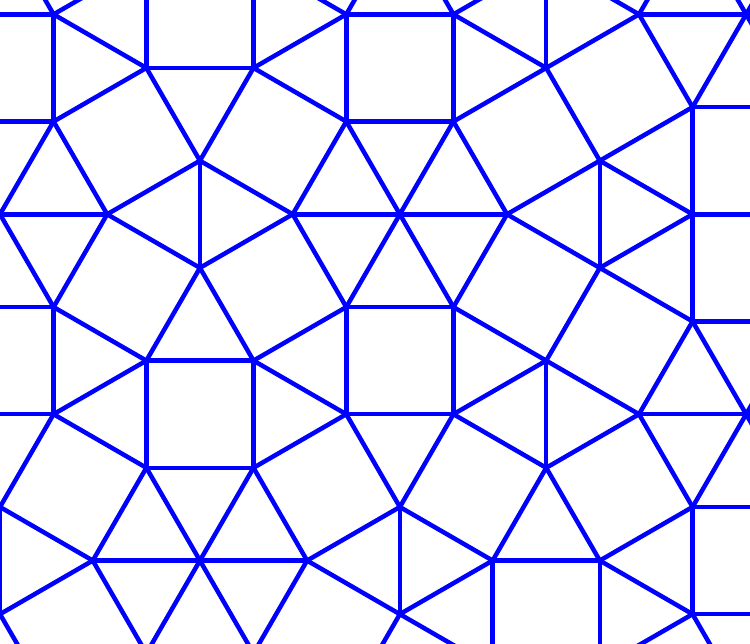}
\caption{(Left) The $\sigma$ lattice. (Right) A portion of the dodecagonal square triangle quasicrystal. }
\label{fig:dodecas}
\end{figure}

The structure we consider has not, to our knowledge, been previously discussed in the literature. We therefore provide some background and additional details about it in this section. As its name implies, the H-$\sigma$ lattice is obtained by decorating the $\sigma$ (or snub-square) lattice with regular hexagons. The parent $\sigma$ lattice is illustrated in Fig.\ref{fig:dodecas}a). It is a member of the family of square-triangle tilings (see ref.(\cite{marianne}) for a classification scheme). It is one of the 11 celebrated Archimedean tilings~\cite{grunbaum} in which all edges are equal, and all vertices are identical modulo reflections.  

\begin{figure}
\centering
\includegraphics[angle=90,width=0.3\linewidth]{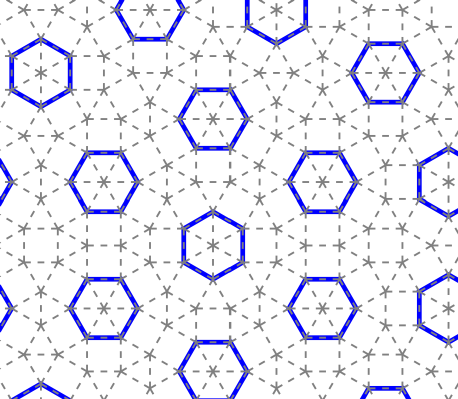}\hskip 1.6cm
\includegraphics[angle=90,width=0.3\linewidth]{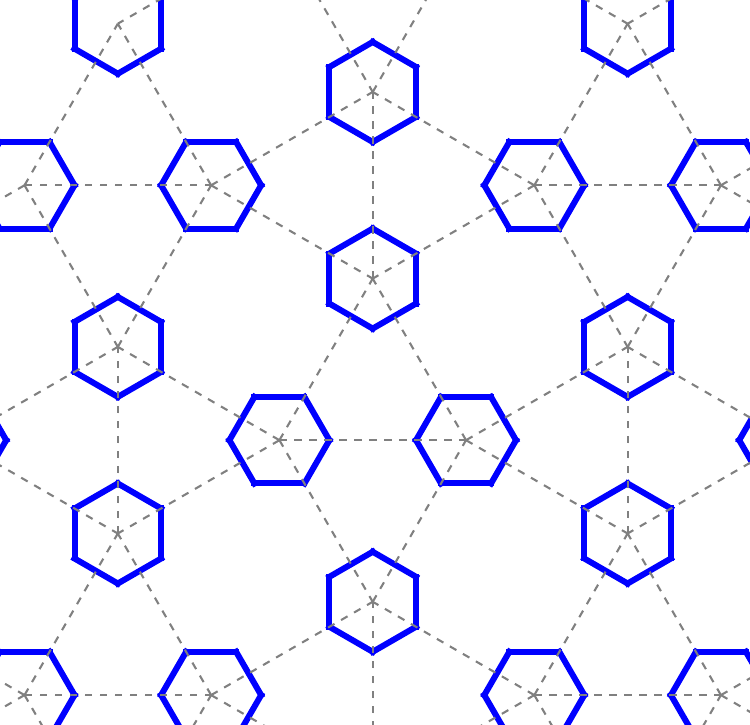}
\caption{(Left) Portion of the quasicrystal showing hexagons highlighted in blue. (Right) The decoration by hexagons of the $\sigma$ lattice using the Schlottman rule. }
\label{fig:sigmalattice}
\end{figure}

There is an infinite variety of square triangle tilings, which include periodic, aperiodic and random tilings. There has been much interest in this family in particular, because it includes 12-fold (dodecagonal) quasicrystals, which are experimentally observed in both soft- and hard- condensed matter systems. Fig.(\ref{fig:dodecas}b) shows a piece of such a dodecagonal quasicrystal made from squares and triangles. If one looks more closely at the quasicrystal, one sees that it contains different types of hexagon, as shown in fig.(\ref{fig:sigmalattice}a) where all of the hexagons have been highlighted in blue. The orientation of these hexagons (some have two vertical edges, while others have two horizontal edges) obeys a rule discovered by Schlottman. In brief, the orientation of each hexagon depends on the symmetry of its local environment (see ref.(\cite{hermisson}) for details).

It turns out that, interestingly, Schlottman's rule which was originally used to generate 12-fold symmetric quasicrystals can also be adapted to other lattices. This is readily done for the $\sigma$ lattice. Using Schlottman's rule one obtains the structure shown in Fig.(\ref{fig:sigmalattice}b). The two different orientations of hexagons appear with equal frequency on the structure. Our model considers spins placed on the vertices of these hexagons and interacting via short range exchange couplings. At the level of nearest neighbor interactions, the system is not frustrated, but becomes so when further neighbor interactions are included, as described in the next section. 

%%%%%%%%%%%%%%%%%%%%%%%%%%%%%%%%%%%%%%%%%%%%%%%%%%%%%%%%%%%%%%%%%%%%%%%%%%%%%%%

To conclude this introductory section, we note that the $\sigma$ lattice is a considerable simplification of the dodecagonal quasicrystal, since it is a periodic lattice, and has a smaller set of local environments. The cluster spin H-$\sigma$ model we introduce in the next section can thus be thought of as a first step towards understanding more complex cluster-spin models for quasicrystalline magnets.

% Our structure is obtained by decorating an Archimedean tiling~\cite{grunbaum} based on squares and triangles, known as the $\sigma$ phase (or snub-square tiling), shown in Fig.(\ref{figsigmalattice}), with hexagonal clusters of six spins. The structure is shown in Fig.(\ref{figHsigmalattice}) along with the square unit cell, outlined in green. Two different orientations of hexagons appear with equal frequency on the structure -- hexagons with vertical sides, and hexagons with horizontal sides. This decoration of the lattice is obtained by adapting the rules given by Schlottman for generating square triangle tilings using an inflation method (see ref.(\onlinecite{hermisson}) for details). 
%%%%%%%%%%%%%%%%%%%%%%%%%%%%%%%%%%%%%%%%%%%%%%%%%%%%%%%%%%%%%%%%%%%%%%%%%%%%%%%%%
\section{Heisenberg model on the H-$\sigma$ lattice}
A simple model for describing the magnetism of localized spins is the
isotropic Heisenberg model. We define it by the following
Hamiltonian~: 
\begin{equation}
H = J_1\sum_{\langle i,j \rangle}\mathbf{S}_i\cdot\mathbf{S}_j + J_2\sum_{[  k,l ]}\mathbf{S}_k 
\cdot \mathbf{S}_l ,
\label{energy}
\end{equation}
where $\langle i,j \rangle$ denotes nearest-neighbor pairs inside hexagons
in red in fig.(\ref{figHsigmalattice}) and $[k,l]$ is the sum over bonds
belonging to triangles in blue in fig.(\ref{figHsigmalattice}). We 
study the classical limit of this Hamiltonian, becoming simply a classical energy.
The spins $\mathbf{S}_i$ are now three-component unit vectors.

This classical spin model can be relevant either to magnetic materials with large spin values 
or temperatures high enough so that quantum fluctuations can be neglected.
As the H-$\sigma$ lattice is a two-dimensional system with full spin rotation symmetry, there is no finite-temperature phase transition. The spin correlation length should increase
with decreasing temperature, diverging only at $T\rightarrow 0$. This does not
preclude transitions breaking discrete symmetries however. For a finite-size
sample we expect that for low enough temperature all spins will adopt
some ground state configuration when the correlation length for magnetic order
is larger than the sample size.

\subsection{Ground state energy}
{ When $J_1<0$ (ferromagnetic nearest neighbor couplings) in Eq.(\ref{energy}), the model is unfrustrated. The ground state is either a simple ferromagnet (for $J_2<0$) or an antiferromagnet formed of clusters of parallel spins. When $J_1 >0$, there is frustration whatever the sign of $J_2$. Indeed, if we assign sites on each type of hexagon to a different sublattice, changing the sign of $J_2$ simply corresponds to changing the sign of spins on one of the sublattices. We therefore confine our attention to the case $J_1, J_2>0$ in the following discussion. }
It is straightforward to obtain an exact
lower bound to the classical energy for the model in Eq.(\ref{energy}). Rewriting the energy Eq.(\ref{energy})
as a sum of squares plus non frustrated bonds, the energy of a single
triangle can be written as~:
\begin{eqnarray}
E_{\vartriangle} &=&J_2\,\mathbf{S}_0\cdotp(\mathbf{S}_1+\mathbf{S}_2)
+J_1\,\mathbf{S}_1\cdotp\mathbf{S}_2 
\label{firstsum}
\\
&=& \frac{1}{2}J_1 (\mathbf{S}_1+\mathbf{S}_2+\frac{J_2}{{J_1}}\mathbf{S}_0)^2 -\left(J_1 +\frac{J_2^2}{2J_1}\right),
\label{secondsum}
\end{eqnarray}
where $\mathbf{S}_0$ is the apex spin common to the two $J_2$ bonds
and the two other spins are noted $\mathbf{S}_{1,2}$.
The complete formula for the energy is now given by~:
\begin{equation}
 E=\sum_{\vartriangle\alpha} E_{\vartriangle}^{(\alpha)}
 + J_1\sum_{i,j}\mathbf{S}_i\cdotp\mathbf{S}_j ,
\end{equation}
where the first sum runs over all the triangles of the lattice and the second sum
involves all the hexagonal bonds not belonging to triangles.

Let us first discuss two simple limiting cases. (i)
When $J_2=0$ one has only decoupled hexagons
and the hexagons are thus N\'eel ordered independently resulting in an extensive
ground state degeneracy for elementary reasons. The ground state energy
is $E_0/N=-J_1$ with $N$ the number of spins.
(ii) One can construct a ferrimagnetic ground state by arranging spins
antiferromagnetically along $J_2$ bonds and ferromagnetically along $J_1$
bonds~: $ \mathbf{S}_{1}=\mathbf{S}_{2}=-\mathbf{S}_{0}$
by using the numbering inside each triangle of formula Eq.(3).
Such a spin configuration can be extended through the entire lattice
and leads to a ferrimagnetic configuration of energy
$E_0/N=-J_1/3-2J_2/3$. This collinear spin configuration has only conventional
magnetic properties.

There is another nontrivial spin configuration that can be constructed
from the energy expression for single triangle.
Let us first search the minimum energy of a single triangle.
First set the square in Eq.(\ref{secondsum}) to zero~:
\begin{equation}
\mathbf{S}_1+\mathbf{S}_2+\frac{J_2}{{J_1}}\mathbf{S}_0 = 0.
\end{equation}
one sees that this condition requires that
the angle $\theta$ between $\mathbf{S}_0$ and $\mathbf{S}_{1,2}$ is given by
\begin{equation}
 \cos\theta = -{J_2}/{2J_1},  
 \label{thetavalue}
\end{equation}
and the angle between $\mathbf{S}_1$ and $\mathbf{S}_{2}$ is $2\pi-2\theta$.
This is feasible as soon as $J_2 \leq 2J_1$.
The triangle energy is then given by $-\left(J_1 +\frac{J_2^2}{2J_1}\right)$.
This solution when it exists gives the absolute minimum of the first term
in Eq.(\ref{secondsum}). If we are able to match all triangles with this configuration and having antiferromagnetic bonds satisfied in the second term of Eq.(\ref{secondsum})
then we have an absolute minimum of the energy
$E_0/N=-J_1-\frac{1}{6}{J_2^2}/{J_1}$.
When it exists this spin configuration is lower in energy than the ferrimagnetic configuration.

We thus find two regimes~: (1) when $J_2 \geq 2J_1$ we have a ferrimagnetic 
collinear state.
(2) when $J_2 < 2J_1$ the spin configuration becomes noncollinear with the angle
$\theta$ defined in Eq.(\ref{thetavalue}) evolves smoothly between
$\pi/2$ for $J_2\rightarrow0$ and $\pi$ for $J_2=2J_1$. It takes the notable
value $2\pi/3$ when $J_2=J_1$. In this regime we note that our analysis
says how to order locally the spins to obtain the absolute minimum energy
but it is not immediately clear how to match individual triangles to cover the 
whole lattice.

We have numerically searched for the ground state configuration by using
the simple algorithm of alignment with the local field. In this scheme 
one starts with some random
initial spin configuration,
one picks a spin at random, compute the local exchange field and then align
the local spin antiparallel to the local field. This procedure is then repeated
many times until convergence is reached. If we start from a fully random
configuration this algorithm is very often stuck in a metastable state so we
start from many random configurations and evolve each of these independently,
typically 512 starting configurations are used and convergence is reached
when the energy does no longer change to machine precision. 
We have also used the Metropolis algorithm for zero temperature.
In this case a random move of the spin also chosen randomly is generated
and the move is accepted only if it leads to a decrease of the energy.
The alignment with the local field in our case is found to be slightly more 
efficient and so is our preferred method.

In the range $0< J_2<2J_1$ the ground state is seen to be non-trivial. For this interval, our simulations show firstly that
the energy always converges to the value given by $E_0$. However, the spin configurations found after reaching convergence are non-collinear and
non-coplanar and have no simple pattern if we
look at them in real space. We find that all angles found numerically
are always given by Eq.(\ref{thetavalue}). Such configurations are thus legitimately
called ground states since they reach the absolute lower bound, $E_0$ found above.
However most importantly we observe that there is definitely a ground state
degeneracy beyond the simple global rotation of all spins
that we describe in the next section.

\subsection{Ground state degeneracy}

\begin{figure}[t]
 \centering
\includegraphics[width=0.8\linewidth]{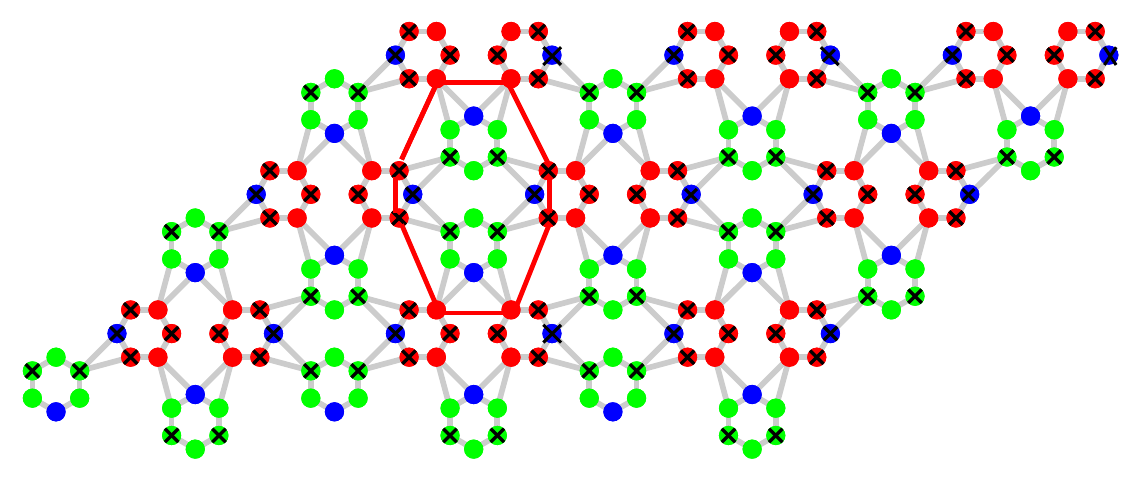} \hskip 1cm 
\hskip 0.5cm
\caption{A ground state spin configuration for a 216 spin sample of the H-$\sigma$ lattice.
Colored dots~: red (A), blue (B), green (C) represent three spin directions such that $A+B+C=0$. Crosses indicates a change of sign. This spin configuration can be periodically continued
in the 2D plane, and corresponds to a non-zero total magnetization  $M=(\sqrt{3}/12)M_{sat}$.
The red polygon is the smallest closed path having spins of only one direction (here $A,\overline{A}$) on the boundary. The spins inside can be arbitrarily rotated with respect to the $A$-axis without changing the energy. Such a ``weathervane'' move
changes the magnetization. }
\label{MaxMag}
\end{figure}

\begin{figure}[t]
\centering
\includegraphics[width=0.8\linewidth]{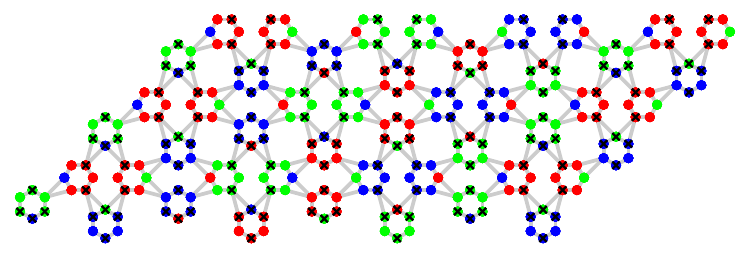} \hskip 1cm 
\hskip 0.5cm
\caption{Another ground state spin configuration for a 288 spin sample 
and $J_2=J_1$
as in Fig.(\ref{MaxMag}). Now the total magnetization is zero for an infinite sample. The configuration can be periodically continued
in the 2D plane.}
\label{ZeroMag}
\end{figure}

To characterize this degeneracy we first make  simple observations in the case
$J_1=J_2$ where all angles are equal to $2\pi/3$ and look for 
coplanar configurations. We note $A, B, C$ three spins with zero sum
that gives a ground state configuration for a triangle.
 Since there are bonds that do not belong to triangles
we  note that
one has to introduce the opposites of $A, B, C$, represented by 
$\overline{A},\overline{B},\overline{C}$. With these six spin directions, it is 
possible to generate
configurations satisfying exactly the lower bound of the energy.
If we try to pave the whole plane it is easy to see that there is indeed
no unique way to do it. We demonstrate this property by exhibiting explicit examples.
In Fig.(\ref{MaxMag}) we display one ground state configuration and
in Fig.(\ref{ZeroMag}) another distinct  ground state configuration. They
do not have the same magnetization. Indeed while the configuration in 
Fig.(\ref{MaxMag})
has $M=(\sqrt{3}/12)M_{sat}$ (where $M_{sat}=N$ is the fully ferromagnetic magnetization value) the configuration in Fig.(\ref{ZeroMag}) has 
zero net magnetization. The finite samples shown in Figs(\ref{MaxMag},\ref{ZeroMag})
can be propagated to infinite lattice sizes.
The two configurations we have displayed suggests that there is a distribution
of possible values of the total magnetization, as we will now discuss.

\subsection{Weathervane modes}
While the order in the H-$\sigma$ lattice has some characteristics in common with
the Kagome Heisenberg antiferromagnet we now show that the situation is quite different.
Indeed while the Kagome system has an extensive degeneracy for the coplanar configurations obtained by permuting $A, B, C$ labels locally, this is not the case
for the H-$\sigma$ lattice. The Kagome lattice Heisenberg model in addition has  
non-planar
ground states. For the H-$\sigma$ lattice too, one can generate nonplanar ground states -- so-called "weathervane" modes~\cite{Ritchey93} which
allow out of plane rotations of the spins at zero energy cost. In the configuration
 shown in Fig.(\ref{MaxMag}) we observe that one can draw a path on the lattice made only
 of say $A,\overline{A}$ spins without traversing any bonds relating to
 $B,\overline{B}, C,\overline{C}$ spins. The path drawn in Fig.(\ref{MaxMag}) is the smallest possible one. It relates only spins of type $A,\overline{A}$ colored in red
 while inside the path one has only $B,\overline{B},C,\overline{C}$ spins 
 colored in blue and green. The path does not intersect any exchange bonds.
 One can then rotate all the spins only
 inside the closed path by an arbitrary angle around the axis defined by spin of type
 $A,\overline{A}$. This generates nonplanar configurations with the same energy, here,
 the ground state energy. These modes lead to
configurations that do not have the same magnetization. This is due to the
 fact that there are spins on each hexagon which do not belong to any triangle, unlike the Kagome lattice.
 We conclude that the ground degeneracy in this lattice also involves  a continuous 
 distribution of the magnetization, in contrast to the Kagom\'e ground states which have all zero
 magnetization. The minimum value of the magnetization per site
 is zero as exemplified by the example in Fig.(\ref{ZeroMag}). The maximum value we find in Monte Carlo studies is $M=\sqrt{3}M_{sat}/12$ of configuration in Fig.(\ref{MaxMag}). 
 We do not however have an explicit proof that this is indeed the maximum value.

%%%%%%%%%%%%%%%%%%%%%%%%%%%%%%%%%%%%%%%%%%%%%%%%%%%%%%%%%%%%%%%%%%%%%%%%%%%%%%% 
\section{Finite temperature properties}
 At finite temperature, it is the entropy that selects
the most probable configuration and determines the macroscopic magnetization in this model. 
In the context of frustrated spin systems this phenomenon
is commonly called ``order by disorder''. We use Monte-Carlo simulations to reproduce
the effect of finite temperature~\cite{Zhito2002,Zhito2008,Davier23,Gembe23,Pitts2022,Richter2024,Schroder2005,Konstantinidis2005,Konstantinidis2007,Konstantinidis2018,Zhito2022,Pinettes2002,Domenge2005,Messio2011,Gvodzikova2016,Domenge2008,Schmidt2005,Schnack2009,Schmidt2003}. Spin configurations are updated by standard
Metropolis steps followed by overrelaxation moves that do not change the energy
but enable better sampling of the configuration space. We perform between two and 
five overrelaxation moves for each Metropolis step. Such a move consists of a 
$\pi$ rotation  
of the spin
around the local exchange field, a deterministic process that requires not much
extra computer time.
The amplitude of the random move of the spins is adjusted to have an acceptance rate
close to 0.5.
Along the Monte Carlo time history we measure the energy as well as the total
magnetization defined as~:
\begin{equation}
 M=\frac{1}{N}\langle|\sum_{i=1}^N \mathbf{S}_i|\rangle
 \label{magdef}
\end{equation}
 We have studied in detail the
magnetization distribution as a function of the temperature in order to understand
the consequences of the many ground states with non fixed magnetization.
System sizes we studied are $24\times L^2$ with $L=4,6,9$ hence $N=384,864,1944$ spins.
To each thermodynamic equilibrium we lower the temperature by cooling the system
to $T=1,0.1,0.01,0.005,0.001 (J_1)$ and checking equilibrium at each intermediate
temperature. All runs are repeated 512 times. A complete cooling run involves
$\approx 10^8$ MC steps. At the lowest temperature the autocorrelation time of the energy
is measured $\approx 10^4$ MC steps and we perform measurements separated by 10 times
this scale.

To interpret the results obtained, let us first discuss what is expected in well-understood cases. A two-dimensional
Heisenberg ferromagnet has no long-range order at any nonzero temperature
due to the Mermin-Wagner theorem. Its spin correlation length grows exponentially
as one decreases the temperature. If we consider a finite piece of lattice
at some low enough temperature the correlation length will exceed the lattice size
and the system will appear to be ordered, with a magnetization close to
the ground state magnetization -- a finite size effect.
The MC measurements of the magnetization define the probability
distribution of the magnetization $P(M)$ which is proportional  
to the histogram of the magnetization values.
If we observe the probability distribution $P(M)$ of the modulus of total magnetization
as a function of temperature, then at low temperature we expect a single peak centered at the saturation value of the finite system and as we increase the temperature
this peak will shift to lower values ultimately reaching the neighborhood of
zero magnetization, when the finite spin system is fully decorrelated.
This is indeed the behavior we observe in the H-$\sigma$ lattice system when $J_2\geqslant 2J_1$. Here $P(M)$ is peaked at the unique  ground state value $M=M_{sat}/3$
and the width of the peak is very small at low temperature.

\begin{figure}
\includegraphics[width=0.4\linewidth]{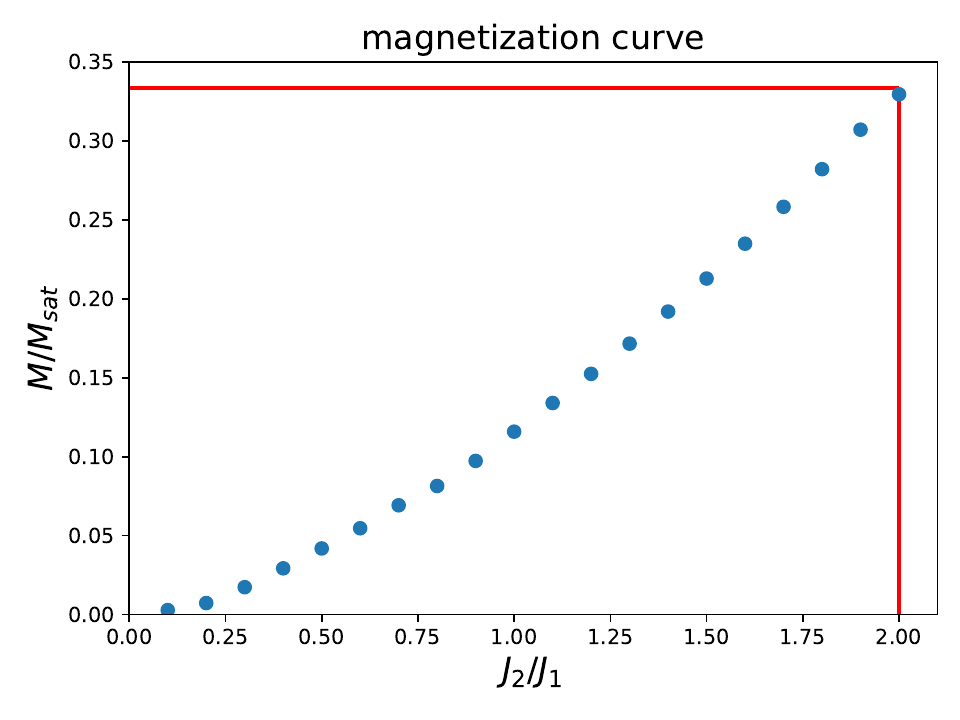} 
\caption{Plot of the average total magnetization $M/M_{sat}$ for a 384-spin sample of the H-$\sigma$ lattice as a function of the coupling ratio $J_2/J_1$. For $J_2 > 2J_1$ the ground state is a simple ferrimagnet with $M/M_{sat}=\frac{1}{3}$ 
(red horizontal line on top of the figure).}
\label{Magcurve}
\end{figure}

\begin{figure}[t]
\centering
\includegraphics[width=0.5\linewidth]{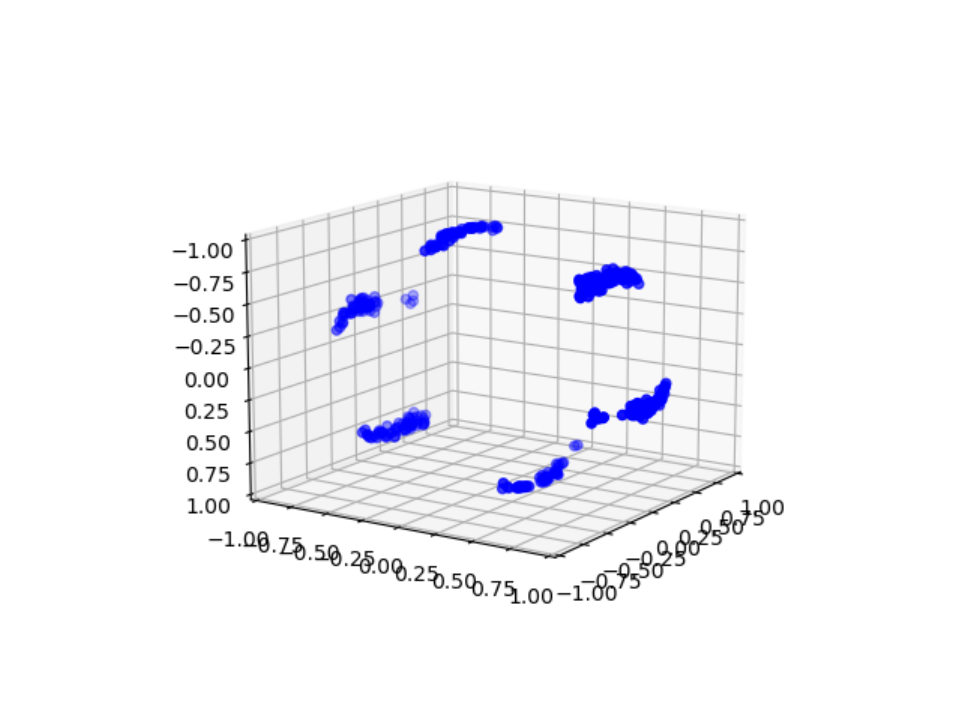} 
\caption{The common-origin plot of the spin configuration with minimum free energy (most probable configuration as deduced from the $P(M)$ histogram) at a temperature $T=10^{-3}J_1$ and $J_2=J_1$, for a sample of 384 spins. 
All the spins (normalized to 1) are plotted with their origin at the origin of the three-dimensional
space. This configuration is not planar
and is not commensurate.}
\label{COOplot}
\end{figure}

If we now turn to the regime $0<J_2< 2J_1$ we find a very different structure. 
\begin{figure}
\centering
\includegraphics[width=0.4\textwidth]{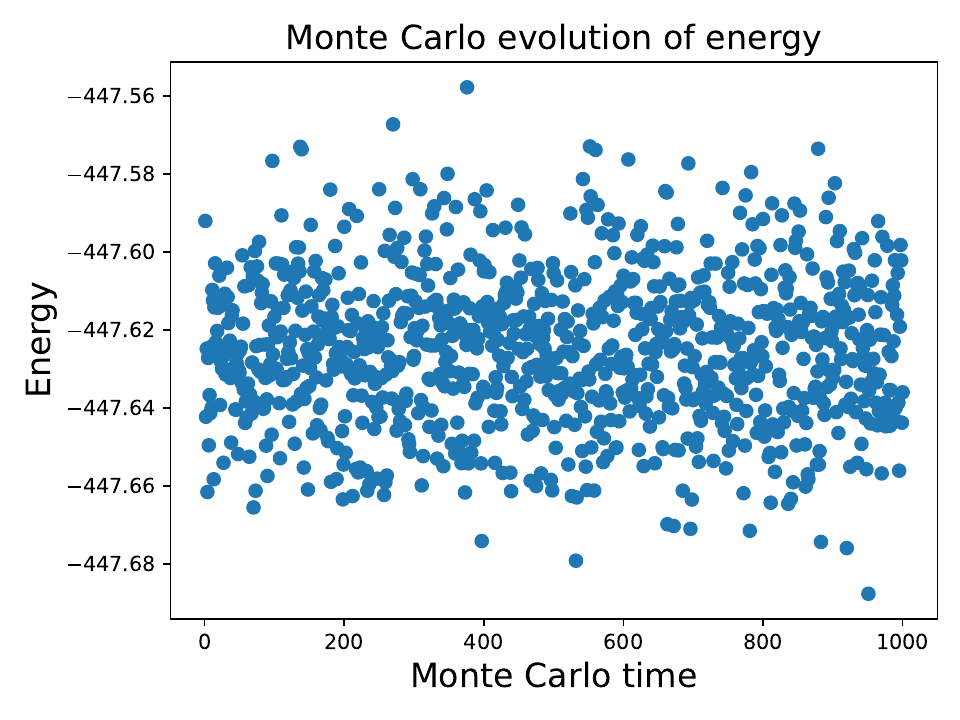}  
\includegraphics[width=0.4\textwidth]{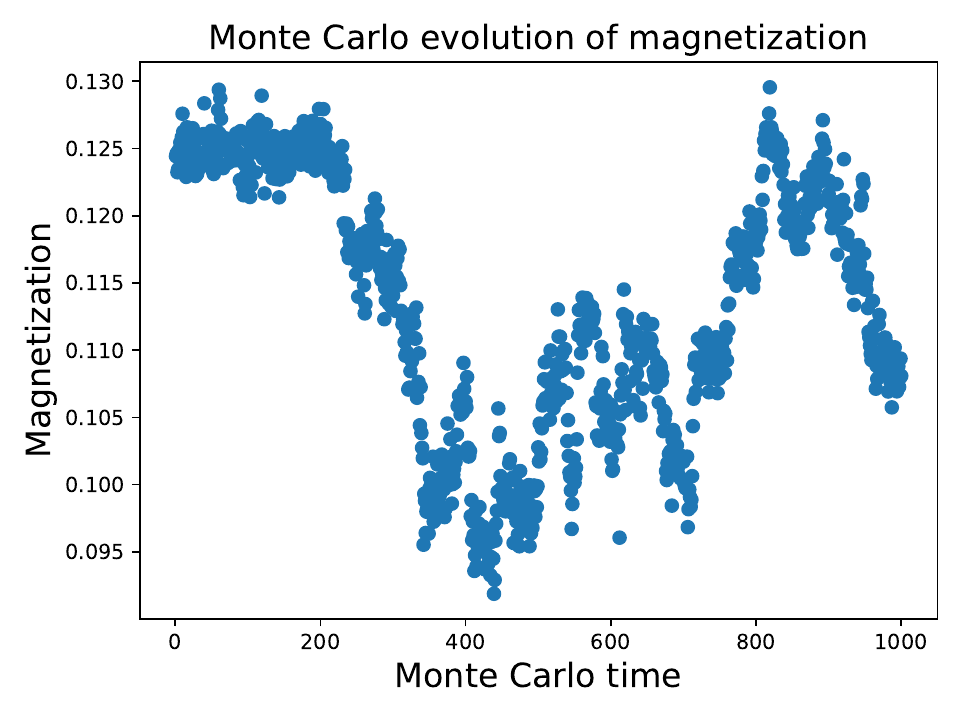} 
\caption{Monte Carlo measurements of energy in units of $J_1$ (left panel) and total magnetization (right panel) from Eq.(\ref{magdef}) as a function of Monte Carlo time steps.
The energy measurements are spaced by ten times the measured autocorrelation time. They leads to a Gaussian histogram centered at a well-defined average energy. 
On the contrary the magnetization
exhibit a complex behavior spending long time in several distinct values.
The system has 384 spins and $J_2=J_1$ and the temperature is $T=10^{-3}J_1$}
\label{mcevolution}
\end{figure}
\begin{figure}
\includegraphics[width=0.5\linewidth]{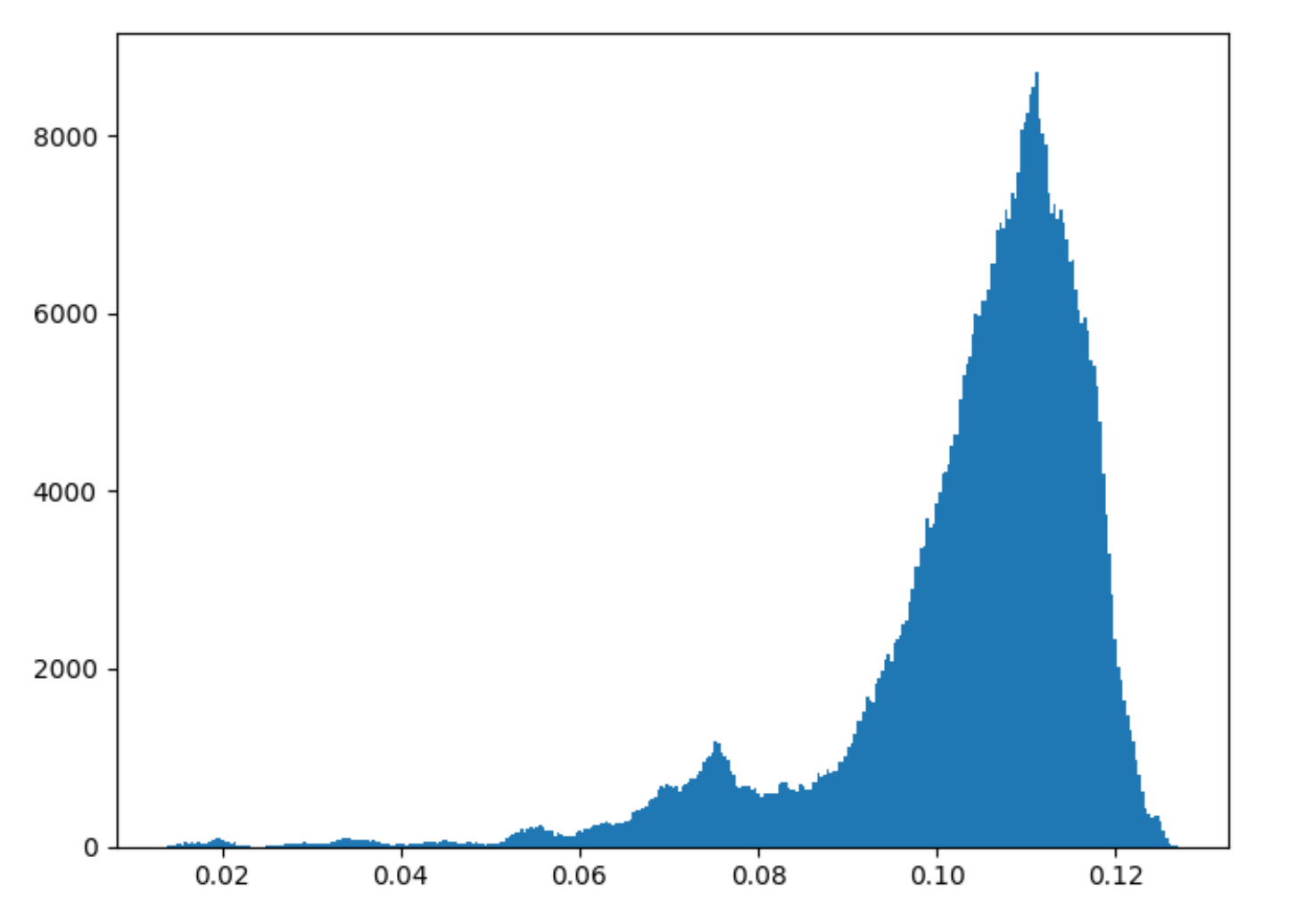}  
\caption{The probability distribution of $M/M_{sat}$ 
obtained by Monte Carlo runs for a 864-site cluster and low temperature, $T=10^{-3}J_1$. 
The distribution extends all the way to zero and has a peak at $M/M_{sat}=0.11$. The fine 
structure seen beyond the main peak is non-random, specific of the lattice.
Multiple runs made with 512 independent replicas always give the same
equilibrium structure.}
\label{Histo1}
\end{figure}
\begin{figure}
\includegraphics[width=0.5\linewidth]{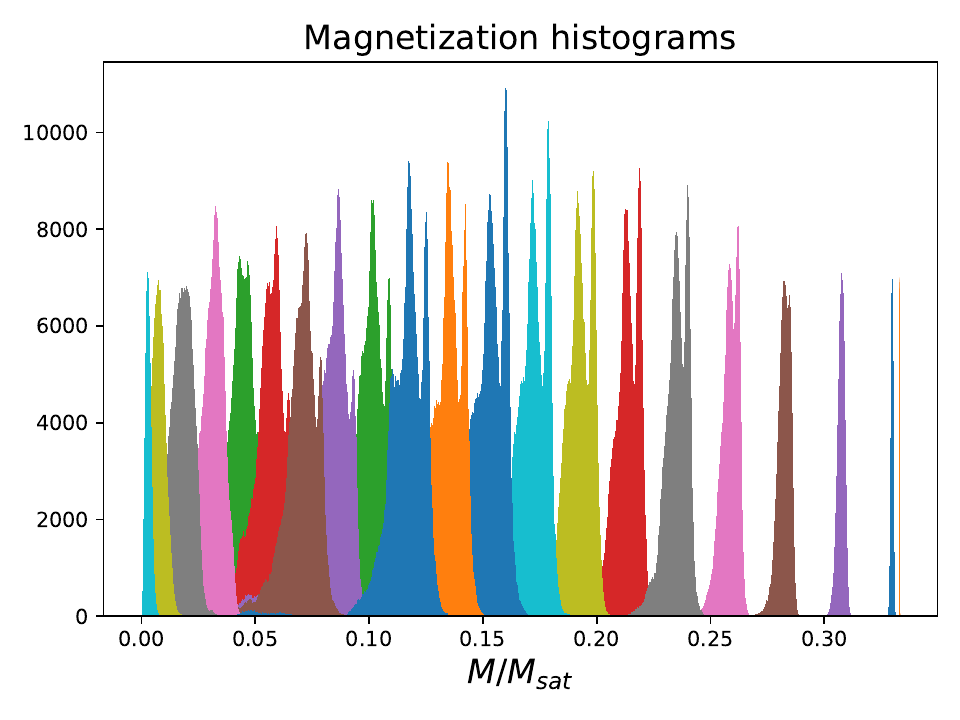} 
\caption{Histograms of $M/M_{sat}$ obtained by Monte Carlo runs.
Each color refers to a different value of the ratio $J_2/J_1$. From left to right $J_2/J_1=0.1$ to $2$ by steps of $0.1$. The rightmost histogram for
$J_2/J_1=2$ corresponds to the ferrimagnetic ground state with no special degeneracy~: this leads to a sharp peak whose width is solely due to finite temperature. For all data the temperature is $T=10^{-3}J_1$. We note that
the distribution of magnetization has a complex shape and in some cases has two
maxima. }
\label{multihistos}
\end{figure}
The function $P(M)$ is now smeared over an extended range going down
to zero magnetization with a peak at some non-trivial magnetization which is
different from the maximum possible value that we find by the explicit construction
given above. In Fig.(\ref{Histo1}) an example of this behavior is displayed
for a 864-spin cluster at $T=10^{-3}J_1$. The peak occurs for $M=0.118M_{sat}$
below the maximum value reached in the configuration of Fig.(\ref{MaxMag}).
The peak value is the magnetization selected by free energy minimization.
This scenario happens for all sizes we have studied and also for the whole
range of exchange couplings $0<J_2\leqslant 2J_1$. In Fig.(\ref{multihistos})
are displayed the histograms of the magnetization for a 384-cluster for values
in this interval. We note that there is collapse for limiting values~:
for $J_2\rightarrow 0$, the limit of decoupled hexagons which are N\'eel ordered
so have zero net magnetization. Similarly the ferrimagnetic phase for 
$J_2\geqslant 2J_1$ has only one sharp peak at the analytically known value of 
$M_{sat}/3$.
Defining the magnetization as the average value, we obtain a magnetization curve as a function of the ratio $J_2/J_1$
displayed in Fig.(\ref{Magcurve}).
Note that with the non-Gaussian special distribution of the magnetization
the average value is never equal to the most probable value
throughout the phase $0<J_2<2J_1$. This phenomenon persists for all lattice sizes we studied.
The spin configuration realized by the minimum of the free energy can be obtained
by selecting the magnetization corresponding to the peak value in the MC process.
The observed state has a complex structure, neither coplanar nor commensurate. This is best seen in a common origin plot, Fig.(\ref{COOplot}) showing the spin vectors (normalized to unity) of all of the sites for a 384 site sample. It can be seen that they are distributed on the unit sphere, indicating that this configuration is non-planar or incommensurate with the lattice or both.

%%%%%%%%%%%%%%%%%%%%%%%%%%%%%%%%%%%%%%%%%%%%%%%%%%%%%%%%%%%%%%%%%%%%%%%%%%%%%%%
\section{external magnetic field effect}
To study the stability of the degenerate phase we apply an external magnetic
field $H$. The energy is given by~:
\begin{equation}
 E = J_1\sum_{\langle i,j \rangle}\mathbf{S}_i\cdot\mathbf{S}_j + J_2\sum_{[  k,l ]}\mathbf{S}_k 
\cdot \mathbf{S}_l -\mathbf{H}\cdot\sum_i \mathbf{S}_i ,
\label{energywithH}
\end{equation}

\begin{figure}
\includegraphics[width=0.4\textwidth]{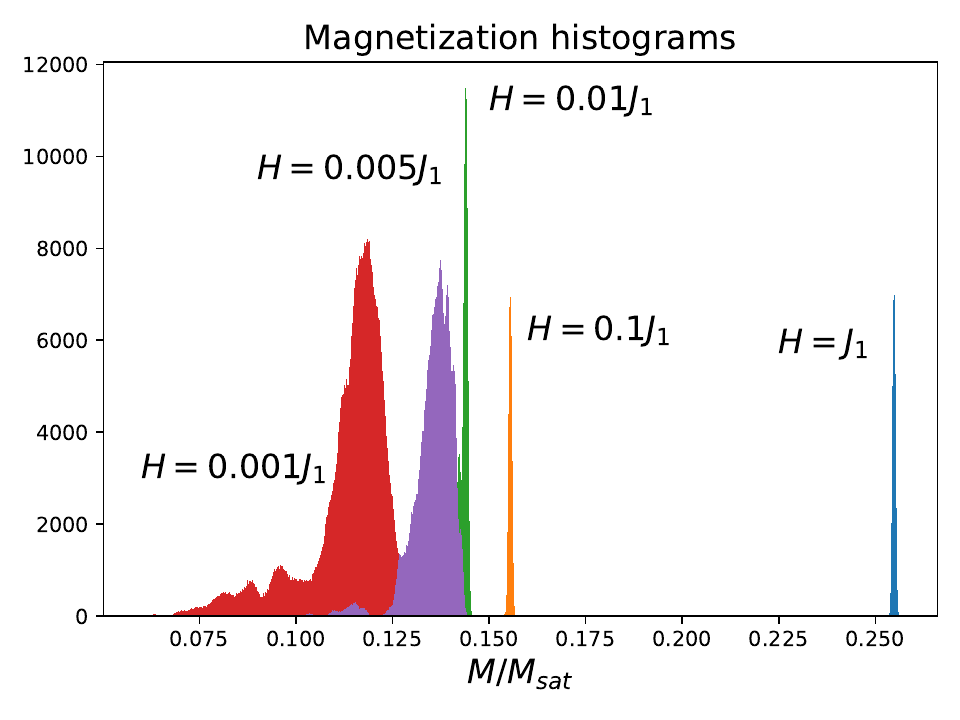} \hskip 1cm 
\hskip 0.5cm
\caption{Histograms of the magnetization at $T=10^{-3}J_1$ for a system with 384
spins and an external magnetic field $H$. We have used the special ratio of
couplings $J_2=J_1$.
The peculiar degeneracy specific to the H-$\sigma$ lattice is destroyed when H
is larger than $\approx0.01 J_1$.
The distribution of magnetization becomes that of a conventional magnet
as observed in the ferrimagnetic phase for $J_2\geqslant 2J_1$.}
\label{histoH}
\end{figure}

To study this situation we use again Monte-Carlo simulations.
For small enough $H$ the ground state degeneracy is still present
with a non-trivial magnetization distribution.
We find that a \textit{finite} value of $H$ leads to lifting of the 
ground state degeneracy.
Indeed the histogram of the magnetization at finite temperature collapses to a single
sharp Gaussian peak only when $H$ is large enough i.e. for $H\gtrsim 0.01J_1$. Several histograms are displayed as a function
of $H$ in Fig.(\ref{histoH}). In the presence of an external field the magnetization
is no longer bounded by the special value $(\sqrt{3}/12)M_{sat}$ of the extremal
configuration in Fig.(\ref{MaxMag})~: the upper bound grows continuously with $H$.
This analysis has been performed only for the special point $J_1=J_2$
but we expect the result to be valid in the whole range
of magnetization degeneracy $0<J_2/J_1<2$. In Figure (\ref{histoH}) the temperature of th simulation is fixed at $T=10^{-3}J_1$.

%%%%%%%%%%%%%%%%%%%%%%%%%%%%%%%%%%%%%%%%%%%%%%%%%%%%%%%%%%%%%%%%%%%%%%%%%%%%%%%%
\section{Longer-range couplings}
\begin{figure}
\includegraphics[width=0.4\textwidth]{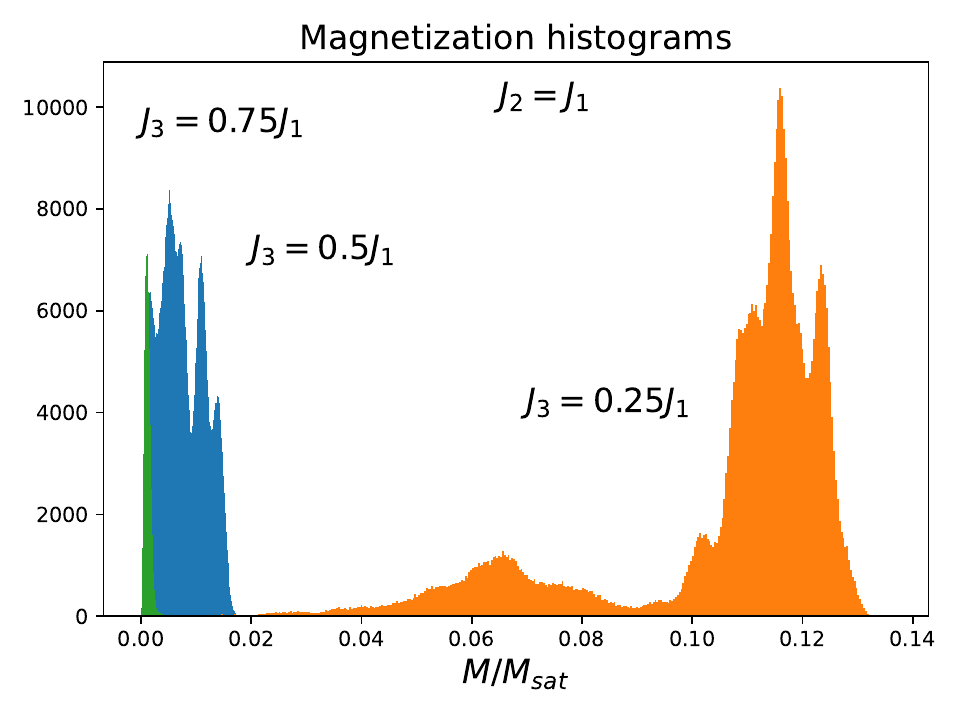} \hskip 1cm 
\hskip 0.5cm
\caption{One may destabilize the ground state degeneracy by adding an extra exchange coupling $J_3$ between the dangling spins between neighboring hexagons.
When this coupling is antiferromagnetic it creates frustration that destroys
the degeneracy provided its strength is beyond $\approx 0.5J_1$.
The system has 384 spins and the temperature is $T=10^{-3}$.}
\label{histoJ3}
\end{figure}

If we consider the spins of the H-$\sigma$ lattice they are all member of
an hexagon and among those of a given hexagon there are only two of them
that are not engaged in a $J_2$ exchange interaction.
If there are real magnets described by the H-$\sigma$ phase then the distance between such 
dangling spins belonging to neighboring hexagons
is not much greater that between other pairs of spins. So the exchange between them
may be sizable. Let us introduce $J_3$ the corresponding exchange coupling. It is
easy to see that
if $J_3$ is ferromagnetic then it does not change the nature of the manifold
of ground states since the dangling spins are always ferromagnetically aligned 
in the ground state configurations as seen in 
Figs.(\ref{MaxMag},\ref{ZeroMag}). So the entropic selection of magnetization still 
operates. This is not the case if $J_3$ is antiferromagnetic~: this frustrates the
degenerate configurations. 
Analytic study is not possible so
we have studied the change of magnetization distribution by
running Monte-Carlo simulations by varying  the ratio $J_3/J_1$  while keeping $J_2=J_1$
for simplicity.
Our results are displayed in Fig.(\ref{histoJ3}).
Histograms of $P(M)$ are computed at the point $J_2=J_1$ for temperature
$T=10^{-3}J_1$ and a sample of 384 spins. We have
chosen a set of values $J_3=0.25,0.5, 0.75 J_1$.
There is again a range of stability for the degenerate phase
which is destroyed when $J_3\gtrsim 0.75J_1$.
This means that the special degenerate phase with entropic selection
is robust beyond the simplest exchange model Eq.(\ref{energy}).
Beyond this critical value the system has zero total magnetization. We have not attempted
a detailed description of the new phase.

%%%%%%%%%%%%%%%%%%%%%%%%%%%%%%%%%%%%%%%%%%%%%%%%%%%%%%%%%%%%%%%%%%%%%%%%%%%%%%%
\section{conclusions}
We have studied a lattice classical Heisenberg model with two exchange couplings
with a complex phase in a special range of parameters $0<J_2<2J_1$.
In this regime
the H-$\sigma$ lattice Heisenberg model has an extensive ground state degeneracy with
continuous distribution of the total magnetization. Thus minimizing the energy does
not uniquely determine the magnetization at zero temperature. At finite T, minimization of free energy leads to a selection of a subset of configurations. In other words, the magnetization is entropically selected and the system shows an order-by-disorder.
This peculiar phenomenon is observed in our Monte-Carlo simulations at nonzero temperatures. The magnetization at finite temperature has a nontrivial probability distribution whose width does not shrink when increasing the lattice size.
The average value of the magnetization does not coincide with the most probable value.
We have obtained the average magnetization of the system in the whole phase diagram.
This very special property deserves more detailed studies. It does not happen in the Kagom\'e spin model.

One can ask about the changes in these states when the original model is perturbed. In particular, one can ask about the effects of additional short range couplings within and between hexagons
the H-$\sigma$ lattice. These couplings lead to increased frustration, due to the appearance of more triangles and rings of five spins. We have checked that, for small values of these added couplings the phases described in this paper are not changed qualitatively. However a more complete analysis of the changes in the phase diagram are left for a future study. 

%%%%%%%%%%%%%%%%%%%%%%%%%%%%%%%%%%%%%%%%%%%%%%%%%%%%%%%%%%%%%%%%%%%%%%%%%%%%%%%
\acknowledgments 
We thank H. T. Diep, G. Misguich, H. Orland 
and P. F. Urbani for useful discussions.   This work was granted access to 
the CCRT  High-Performance Computing (HPC) facility under the Grant CCRT2024 
awarded by the Fundamental Research Division (DRF) of CEA.

\end{document}